\documentclass[useAMS,usenatbib]{mn2e}

\usepackage{hyperref}
\usepackage{mathtools}
\usepackage{amsmath,amssymb,epsfig,bm}
\usepackage [english] {babel}
\usepackage[utf8x]{inputenc}
\usepackage{amsfonts}
\usepackage{./aas_macros}
\usepackage{xcolor}
\usepackage{caption}
\usepackage{listings}
\usepackage{placeins}
\usepackage{graphics}
\usepackage{graphicx}
\usepackage{etoolbox}
\usepackage{natbib}
\usepackage{psfrag}
 \newcommand{\mbf}{\mathbf}

\title[Superconducting NS with entrainment]{Magnetized neutron stars with superconducting cores: Effect of entrainment}
\author[K. Palapanidis, N. Stergioulas and, S. K. Lander ]{K. Palapanidis$^{1}$\thanks{E-mail:
kp3g14@soton.ac.uk}, N. Stergioulas$^{2}$ and, S. K. Lander$^{1}$\\
$^{1}$School of Mathematics, University of Southampton, Southampton SO17 1BJ, U.K.\\
$^{2}$Department of Physics, Aristotle University of Thessaloniki, Thessaloniki 54124, Greece
}
\begin{document}

\date{Accepted 0000 December 00. Received 0000 December 00; in original form 0000 October 00}

\pagerange{\pageref{firstpage}--\pageref{lastpage}} \pubyear{0000}

\maketitle

\label{firstpage}

\begin{abstract} We construct equilibrium configurations of
  magnetized, two-fluid neutron stars using an iterative numerical
  method. Working in Newtonian framework we assume that the neutron star has two regions: the
  core, which is modelled as a two-component fluid consisting of
  type-II superconducting protons and superfluid neutrons, and the
  crust, a region composed of normal matter. Taking a new step
    towards more complete equilibrium models, we include the effect of
    entrainment, which implies that a magnetic force acts on neutrons,
  too. We consider purely poloidal field cases and present
    improvements to an earlier numerical scheme for solving
    equilibrium equations, by introducing new convergence criteria. We find that entrainment results in qualitative differences in the structure of field lines along the magnetic axis.
\end{abstract}


\section{Introduction}

Neutron stars are ideal astrophysical laboratories.  Their extreme properties demand an interdisciplinary approach, while most of the existing models attempt to synthesize the various features that neutron stars are predicted to possess.\\
Of particular interest is the construction of numerical solutions for equilibrium configurations of magnetized neutron stars with realistic assumptions about their interior structure.  The Hachisu self-consistent field method (HSCF) \citep{Hachisu1} is the starting point of many modern methods for constructing (initially non-magnetized) equilibria.  This method was extended by \cite{TomimuraEriguchi} so that magnetic fields with a strong poloidal and a weak toroidal component were included.  \cite{Lander2012, Lander2ndA} and \cite{Lander2ndB} further extended the methods, using a two-fluid description, including  type-II superconductivity for the neutron star core. In these latter models, the star is divided in two regions, the superconducting core and the normal infinitely conducting crust. 

Here, we also take into account, for the first time in such numerical models,  the effect of entrainment. This comes by assuming an interaction between the momenta of the two fluids, resulting in an additional force acting on neutrons, see \cite{SCMHD} and references therein. 

Furthermore, because the number of equations to be solved has increased, we show that the    simple convergence criteria introduced in   \cite{Hachisu1} are no longer adequate. We present new convergence criteria, related to  fundamental quantities of the model and show how one obtains more accurate numerical solutions.   

Although our code is built in a way that it is possible to numerically evaluate both purely poloidal and mixed poloidal-toroidal equilibrium models, we focus here on the case of purely poloidal fields. The reason for doing so is that in these models $B_{\phi}\ll B_{\rm pol}$. We compute models with entrainment for a magnetic field strength of  $\sim 10^{15} \rm G$ at the pole and compare to corresponding models without entrainment, that were previously obtained in \cite{Lander2ndB}.

In the following sections we first briefly discuss entrainment, then
we describe the model. We demonstrate our new convergence criteria
along with the numerical method and finally we provide results for
equilibria with and without entrainment, and compare these.

\section{Entrainment in Neutron Stars}

It is theorized that neutron stars exhibit entrainment, a phenomenon occurring in the case of mixed fluids. Due to entrainment, the momenta of the components are coupled, which means that the momentum of each component is dependent on both fluid velocities. Thus, if a component is flowing, the other component will flow as well. Following \cite{PrixComerAndersson} and \cite{SCMHD} the entrainment can be simply described through a dimensionless function $\varepsilon_\star$ defined by
 \begin{eqnarray}
 \varepsilon_\star=\frac{1-\varepsilon_{\rm n}-\varepsilon_{\rm p}}{1-\varepsilon_{\rm n}},
 \label{estar}
 \end{eqnarray}
 where $\varepsilon_{\rm n}$ and $\varepsilon_{\rm p}$ are \textit{entrainment coefficients}, satisfying, by definition, the relation  $\rho_{\rm p}\varepsilon_{\rm p}= \rho_{\rm n}\varepsilon_{\rm n}$, where $\rho_{\rm p}$ and $\rho_{\rm n}$ are the proton and neutron mass densities. We here consider the slow-rotation approximation, neglecting rotation related terms in the equations of hydrostationary equilibrium as the shape of the star is still spherical.\color{black} The coefficient $\varepsilon_{\rm p}$ can be written in terms of the bare proton mass $m_{\rm p}$ and the effective mass $m_{\rm p}^\star$ acquired by each proton as a result of the entrainment, as
  \begin{eqnarray}
 \varepsilon_{\rm p}= 1-\frac{m_{\rm p}^{\star}}{m_{\rm p}},
 \label{ep}
 \end{eqnarray}
 and thus $\varepsilon_\star$ can be written as 
 \begin{eqnarray}
\varepsilon_\star= \frac{\rho_{\rm p}\left(\delta_\star-1 \right)+\rho_{\rm n}\delta_\star }{\rho_{\rm p}\left(\delta_\star-1 \right)+\rho_{\rm n}} ,
 \label{estar2}
 \end{eqnarray}
where we have set $\delta_\star := \frac{m_{\rm p}^{\star}}{m_{\rm p}} $, which depends on the density and thus on the equation of state. In general, $\delta_\star$ varies depending on the model taken into account. Some of these models are discussed in \cite{Chamel}. Here, we assume that it is constant throughout the core. Notice that in the case that $m_{\rm p}^{\star}=m_{\rm p}$,  $\varepsilon_\star=1$ and the entrainment effect vanishes. 

Entrainment allows neutrons to experience a magnetic force. Normally, this force would be absent since neutrons are neutral and thus do not interact with the magnetic field. However, since the protons are assumed to be superconducting and the neutrons superfluid, the drag between the proton and neutron fluid components  provides a mechanism of interaction between neutrons and the magnetic field, which is enacted through the entrainment force on neutrons \citep{Alpar}.

\section{The model}
 
  \subsection{General description}

Working in Newtonian framework we assume a magnetized, axisymmetric neutron star in a stationary state. The axis of symmetry is parallel to the magnetic dipole axis. The star is assumed to have only two regions: the core and the crust. The core, a region that starts at the center and extends to roughly 90\% of the radius of the star (the crust-core boundary), is assumed to consist of superfluid neutrons and type-II superconducting protons, where the protons are subject to a magnetic force different from the normal ideal MHD case,\footnote{For simplicity, we ignore the presence of electrons, as well as more exotic particle species that may appear at high densities in the core.} while neutrons will also be subject to a magnetic force due to entrainment. The crust, which lies between the crust-core boundary and the surface of the star, is composed of normal matter,  which is subject to the standard Lorentz force. The crust is assumed  to be in a relaxed state, without tangential stresses and can thus be described as a single fluid. Outside the star we assume vacuum.
For purely numerical convenience we adopt the crust model described in \cite{Lander2ndA}, \cite{Lander2ndB} and \cite{Lander2012}. In the future, using realistic equations of state, it should be possible to adopt more realistic proton and neutron fractions throughout the star. However, for evaluating the entrainment effects in the core, the assumptions made for the crust are not crucial.

We express the main equations in cylindrical coordinates $(\varpi, \phi, z),$ while in the numerical implementation we use spherical polar coordinates $(r,\theta)$.  Vector components are expressed with respect to the unit vectors $(\mbf{e}_\varpi, \mbf{e}_\phi,\mbf{e}_z)$ while we provide some results in both dimensionless form and Gauss-CGS units. 
 
\subsection{Two-fluid description}  \label{2fluiddescr}
 
In the core, we need to consider two separate equations for the hydrostationary equilibrium \citep[obtained by setting the neutron and proton velocities equal to zero in the general MHD equations presented in][]{SCMHD}:
\begin{eqnarray}
  \bm{\nabla} \tilde{\mu}_{\rm n} + \bm{\nabla} \Phi_{\rm g} = \frac{\mbf{F}_{\rm n}}{\rho_{\rm n}},
 \label{ScEulLanN}
 \end{eqnarray}
 and
  \begin{eqnarray}
  \bm{\nabla} \tilde{\mu}_{\rm p} + \bm{\nabla} \Phi_{\rm g}= \frac{\mbf{F}_{\rm mag}}{\rho_{\rm p}},
 \label{ScEulLanP}
 \end{eqnarray}
 where $\tilde{\mu}_{\rm n}, \tilde{\mu}_{\rm p}$ are the neutron and proton chemical potentials and $\Phi_{\rm g}$ is the gravitational potential \citep[see][]{Mendell}. $\mbf{F}_{\rm mag}$ is the magnetic force acting on protons and $\mbf{F}_{\rm n}$ is the magnetic force on neutrons due to entrainment. We will discuss chemical potentials, as well as the form of the magnetic force in the following sections.

The gravitational field is related to the total density $\rho$ through
  \begin{eqnarray}
  \bm{\nabla}^2 \Phi_{\rm g} = 4\pi G \rho= 4 \pi G (\rho_{\rm n} +\rho_{\rm p}).
 \label{ScPhipot}
 \end{eqnarray}
We subtract (\ref{ScEulLanN}) from (\ref{ScEulLanP}) to find 
  \begin{eqnarray}
  \bm{\nabla} ( \tilde{\mu}_{\rm p} - \tilde{\mu}_{\rm n} )= \frac{\mbf{F}_{\rm mag}}{\rho_{\rm p}} - \frac{\mbf{F}_{\rm n}}{\rho_{\rm n}},
 \label{ScEulLanDif}
 \end{eqnarray}
 which we use along with (\ref{ScEulLanP}) (instead of using (\ref{ScEulLanP}) and (\ref{ScEulLanN})).
 Since all quantities on the left side of (\ref{ScEulLanN}) and (\ref{ScEulLanP}) are gradients of scalars, the same must hold for the right side, so
 \begin{eqnarray}
  \frac{\mbf{F}_{\rm mag}}{\rho_{\rm p}} = \bm{\nabla} M ,
 \label{ScnablaMeqF}
 \end{eqnarray}
 and
\begin{eqnarray}
 \frac{{\bf F}_{\rm n}}{\rho_{\rm n}}={\bm \nabla}  N,
 \label{ScnablaNeqFn}
 \end{eqnarray} 
 where $M$, $N$ are scalar functions. This result is useful, since  (\ref{ScEulLanP},\ref{ScEulLanDif}) then admit first integrals
\begin{eqnarray}
   \tilde{\mu}_{\rm p} + \Phi_{\rm g}  -M -C_{\rm p} =0 ,
 \label{ScIntP}
 \end{eqnarray} 
 and
  \begin{eqnarray}
   \tilde{\mu}_{\rm p} - \tilde{\mu}_{\rm n} -M +N -C_{\rm dif} =0, 
 \label{ScIntDif}
 \end{eqnarray}  
 where $C_{\rm p}$ and $C_{\rm dif}$ are some integration constants. The above first integrals will be used in the numerical implementation.
  The  magnetic field $\mbf{B}$ needs to satisfy the divergence free constraint 
   \begin{eqnarray}
  \bm{\nabla} \cdot \mbf{B} = 0,
 \label{ScnablaB}
 \end{eqnarray}
 which implies that in axisymmetry it can be decomposed as 
 \begin{eqnarray}
  \mbf{B} = \frac{1}{\varpi}\bm{\nabla} u \times \mbf{e}_\phi + B_\phi \mbf{e}_\phi,
 \label{ScBgen}
 \end{eqnarray}
 where $u$ is the streamfunction describing the magnetic field.
 The system is closed with an equation of state in terms of an energy functional, which will be discussed in the next section.  
 
     \subsection{Equation of state}
 
In the two-fluid description, equations (\ref{ScEulLanN}), (\ref{ScEulLanP}) are formulated in terms of the chemical potentials $\tilde{\mu}_{\rm p},\tilde{\mu}_{\rm n}$ instead of the pressure $P$. We assume an equation of state in terms of the \textit{energy functional} $\mathcal{E}(\rho_{\rm n} , \rho_{\rm p})$ given by
     \begin{eqnarray}
   \mathcal{E}= k_{\rm n} \rho_{\rm n}^{1+\frac{1}{N_{\rm n}}} +k_{\rm p} \rho_{\rm p}^{1+\frac{1}{N_{\rm p}}},
 \label{ScEOS}
 \end{eqnarray}
      where $k_{\rm n}, k_{\rm p}$ are constants, and $N_{\rm n}, N_{\rm p}$ are the polytropic indices for neutrons and protons respectively. This is essentially an extension of the usual polytropic equation of state. The chemical potentials are defined through
  \begin{subequations}
\begin{eqnarray} 
   \tilde{\mu}_{\rm p} \equiv \left. \frac{\partial \mathcal{E} }{\partial \rho_{\rm p}} \right|_{\rho_{\rm n}}=k_{\rm n} \left( 1+\frac{1}{N_{\rm n}} \right)\rho_{\rm n}^{\frac{1}{N_{\rm n}}} ,
  \label{Scchpotdef1}
  \end{eqnarray}
  and
  \begin{eqnarray} 
   \tilde{\mu}_{\rm n} \equiv \left. \frac{\partial \mathcal{E} }{\partial \rho_{\rm n}} \right|_{\rho_{\rm p}}=k_{\rm p} \left( 1+\frac{1}{N_{\rm p}} \right)\rho_{\rm p}^{\frac{1}{N_{\rm p}}} .
\label{Scchpotdef2} 
  \end{eqnarray}
 \end{subequations}
 In the limit of a single-fluid description, one recovers the usual equations in terms of pressure, by  
  \begin{eqnarray}
   \bm{\nabla} P = \rho_{\rm n} \bm{\nabla} \tilde{\mu}_{\rm n} + \rho_{\rm p} \bm{\nabla} \tilde{\mu}_{\rm p}.
 \label{ScChpotPres}
 \end{eqnarray}

   \subsection{Type-II superconducting core} \label{SuperconCore}
     
   We assume type-II superconductivity for the protons in the core. Thus, the magnetic force is no longer the familiar Lorentz force, but it is replaced by a flux tube tension force  \citep{SCMHD}
    \begin{eqnarray} 
   \mbf{F}_{\rm mag}= -\frac{1}{4\pi} \left[\mbf{B}\times \left( \bm{\nabla} \times \mbf{H}_{\rm c1} \right) + \rho_{\rm p} \bm{\nabla} \left( B \frac{\partial H_{\rm c1}}{\partial \rho_{\rm p}} \right)  \right] ,
  \label{ScMagForce}
  \end{eqnarray}
  where $\mbf{H}_{\rm c1}=\mbf{H}_{\rm c1}(\rho_{\rm p}, \rho_{\rm n})$ is the first critical field  given by $\mbf{H}_{\rm c1}= H_{\rm c1}\hat{\mbf{B}}$ and $\hat{\mbf{B}}$ is the unit tangent vector to the magnetic field ($\hat{\mbf{B}}= \mbf{B}/B=\hat{B}_\varpi \mbf{e}_\varpi +\hat{B}_\phi \mbf{e}_\phi +\hat{B}_z \mbf{e}_z$), with  $B$ the norm of the magnetic field.
 The norm of the first critical field is approximated by
     
 \begin{eqnarray}
  H_{\rm c1}(\rho_{\rm p},\rho_{\rm n}) = h_{\rm c} \frac{\rho_{\rm p}}{\varepsilon_\star},
 \label{ScHc1}
 \end{eqnarray}    
 where $h_{\rm c}$ is some constant and $\varepsilon_\star$ is the entrainment function (\ref{estar2}).
The neutron force $\mbf{F}_{\rm n}$ is given by
  \begin{eqnarray}
   \mbf{F}_{\rm n}= -\frac{\rho_{\rm n}}{4\,\pi}\bm{\nabla}\left( B\frac{\partial H_{\rm c1}}{\partial \rho_{\rm n}} \right).
 \label{Fn}
 \end{eqnarray}
 This force on neutrons exists due to the coupling of proton and neutron flows at the vicinity of each fluxtube. These flows exist on a mesoscopic level and are different from the macroscopic fluid velocities which enter the fluid motion equations \citep[see][]{SCMHD}. We define $D_{\rm p}$ and $D_{\rm n}$ as
\begin{subequations}
\begin{eqnarray}
   D_{\rm p}:=\frac{\partial H_{\rm c1}}{\partial \rho_{\rm p}},
 \label{Dp}
\end{eqnarray}
and
\begin{eqnarray}
   D_{\rm n}:=\frac{\partial H_{\rm c1}}{\partial \rho_{\rm n}}.
 \label{Dn}
\end{eqnarray}
\end{subequations} 
 If $\varepsilon_\star$ is constant, then $H_{\rm c1}=H_{\rm c1}(\rho_{\rm p})$ and hence $D_{\rm n}=0$. In that case, the force acting on neutrons vanishes. Combining (\ref{ScnablaNeqFn}) and (\ref{Fn}) it is obvious that
     \begin{eqnarray}
  N = - \frac{BD_{\rm n}}{4\pi}. 
 \label{Nquant}
 \end{eqnarray} 
We extend the derivation for the type-II superconducting equivalent of the Grad-Shafranov equation  for the magnetic field  \citep{Lander2ndB} in the case that $H_{\rm c1}=H_{\rm c1}(\rho_{\rm p},\rho_{\rm n})$ 
  \begin{eqnarray}
  \Delta_{\star} u = \frac{\bm{\nabla} \Pi \cdot \bm{\nabla} u}{\Pi}  - \varpi^2 \rho_{\rm p} \Pi \frac{\rm d \it y}{\rm d \it u} - \Pi^2 f_{\rm sc} \frac{\rm d \it f_{\rm sc}}{\rm d \it u},
 \label{ScGS}
 \end{eqnarray}
 where
 \begin{eqnarray}
   \Delta_{\star} u \equiv \left( \frac{\partial^2 u}{\partial\varpi^2} - \frac{1}{\varpi}\frac{\partial u}{\partial \varpi}   +\frac{\partial^2 u}{\partial z^2} \right) \equiv \frac{\varpi}{\sin \phi} \bm{\nabla}^2 \left(  \frac{u \sin{\phi} }{\varpi} \right),
\label{deltaoper}
  \end{eqnarray}
while functions $f_{\rm sc}$ and $y(u)$ are defined through
 \begin{eqnarray}
   y(u):=4\pi M_{\rm sc} +BD_{\rm p},
 \label{yfun}
 \end{eqnarray}
  \begin{eqnarray}
   f_{\rm sc}(u):=\varpi \frac{B_\phi}{B} H_{\rm c1}.
 \label{ffun}
 \end{eqnarray}
 We note here that the contribution of the neutron force $\mbf{F}_{\rm n}$ to the Grad-Shafranov equation is implicit. $B$ is related to $u$ through
  \begin{eqnarray}
 B \equiv  \sqrt{\mbf{B} \cdot \mbf{B} } &=H_{\rm c1} \frac{\left| \bm{\nabla} u \right|}{\sqrt{H_{\rm c1}^2 \varpi^2 -f_{\rm sc}^2}},
 \label{ScBscon}
 \end{eqnarray}
 where we have set
 \begin{eqnarray}
\Pi := \frac{B}{H_{\rm c1}} = \frac{\left| \bm{\nabla}u \right|}{\sqrt{\varpi^2 H_{\rm c1}^2-f_{\rm sc}^2}}.
 \label{Pifun}
 \end{eqnarray}
 We have denoted $M$ and $f$ with subscript sc in order to distinguish
 them from their normal matter counterparts. The derivation of the
 Grad-Shafranov equation (\ref{ScGS})  is shown in section \ref{APgsderiv}.

\subsection{Normal matter crust}
 In this section we discuss the form for the $\mbf{B}$ field in the crust region. There, we assume normal, perfectly conducting matter and hence, the governing equations are those of the ideal MHD used in the first part. The magnetic force is
    \begin{eqnarray}
 \mbf{F}_{\rm mag} = \frac{1}{4\pi} \left( \bm{\nabla} \times \mbf{B} \right) \times \mbf{B},
 \label{ScFmagnormal}
 \end{eqnarray} 
 while the neutron force vanishes
   \begin{eqnarray}
 \mbf{F}_{\rm n} = 0.
 \label{ScFN0}
 \end{eqnarray} 
 The Grad-Shafranov equation for this case is
  \begin{eqnarray}
  \Delta_{\star} u = -4\pi \varpi^2 \rho_{\rm p} \frac{\rm d \it M_{\rm N}}{\rm d \it u} - f_{\rm N} \frac{\rm d \it f_{\rm N}}{\rm d \it u},
 \label{ScGsnormal}
 \end{eqnarray} 
       where we have denoted the functions $M,f$ with subscript N to distinguish them from their superconducting counterparts. It can be shown that both $M_{\rm N}$ and $f_{\rm N}$ are functions of $u$. As in the superconducting case, $f_{\rm N}$ is related with the toroidal part of the magnetic field $B_\phi$ through
   \begin{eqnarray}
 f_{\rm N}(u) = \varpi B_\phi .
 \label{fuN}
 \end{eqnarray} 
       
   \subsection{The exterior}
   The exterior of the star is assumed to be a perfect vacuum, since we have not assumed the presence of a magnetosphere. Therefore, the matter density vanishes in the outer region and the equation governing the magnetic field is
  \begin{eqnarray}
  \Delta_{\star} u = 0.
 \label{ScGsvac}
 \end{eqnarray}      
 It is possible to model a magnetosphere by assuming that part of the toroidal component of the magnetic field exceeds the surface of the star, as presented in \cite{magnetosphere}.
\subsection{Boundaries and boundary conditions}
 
For specifying the various regions of the neutron star, as well as the boundary conditions, we use the same assumptions as in \cite{Lander2ndB}, modified in a way that entrainment is included.  
 
 \subsubsection{The crust-core and surface boundaries} 
 
 The surface of the neutron star is defined by the contour of vanishing proton density 
  \begin{eqnarray}
  \rho_{\rm p}^{\rm surf}(\varpi,z) = 0 ,
 \label{NSsurf}
 \end{eqnarray} with a corresponding radius $r_e:=r^{\rm p}_{\rm
eq}$ in the equatorial plane. The core of the neutron star extends from the center out  to the proton density contour that is at a  radius $0.9\,r^{\rm p}_{\rm eq}$ in the equatorial plane, which also defines the crust-core boundary 
 \begin{eqnarray}
  \rho_{\rm p}^{\rm cc} (\varpi, z) := \rho_{\rm p} (0.9\, r^{\rm p}_{\rm
eq},0). 
 \label{ccsurf}
 \end{eqnarray} 
The crust is the region extending between the above crust-core boundary and the surface of the star.
 
 \subsubsection{The boundary conditions}
 
   The first boundary condition to be met is the continuity of the magnetic force on the crust-core boundary (we denote the crust-core boundary using the cc subscript)
 \begin{eqnarray}
 \left[ \rho_{\rm p}^{\rm core} \bm{\nabla} M_{\rm sc} \right]_{\rm cc} = \left[ \rho_{\rm p}^{\rm  crust}\bm{\nabla} M_{\rm N}\right]_{\rm cc}.
 \label{Scbound1}
 \end{eqnarray}      
 Substituting (\ref{yfun}) in (\ref{Scbound1}) and assuming that $M_{\rm N}$ is a function of $u$ only, we obtain (see \ref{APboundcond})
 \begin{eqnarray}
    \left[ \bm{\nabla}\left( B D_{\rm p} \right) \right]_{\rm cc} = \left[ \left( \frac{\rm d \it y}{\rm d \it u} - 4\pi  \frac{\rho_{\rm p}^{\rm  crust}}{\rho_{\rm p}^{\rm  core}}  \frac{\rm d \it M_{\rm N}}{\rm d \it u} \right) \bm{\nabla} u\right]_{\rm cc}.
 \label{Scbound1a}
 \end{eqnarray}      
 It is obvious that $B_{\rm cc}=B_{\rm cc}(u)$. Since we do not know $B$ as a function of $u$ on the core-crust boundary explicitly, we will use a polynomial approximation for $B_{\rm cc}$, denoted by $\tilde{B}_{\rm cc}(u)$, employing the exact same scheme as in \cite{Lander2ndB}
  \begin{eqnarray}
\tilde{B}_{\rm cc}(u)= c_0 +c_1 u + c_2 u ( u- u_{\rm cc}^{\rm eq}),
 \label{ScBtilde}
 \end{eqnarray}      
 where $c_0,c_1,c_2$ are constants and $u_{\rm cc}^{\rm eq}$ is the equatorial value of $u$ on the crust-core boundary. The constants are chosen in such a way that the polynomial values coincide with the numerical ones at the pole and equator. Since we chose a second degree polynomial, a third point is needed as well, which we choose to be at the middle of the $\theta$ direction ($\theta = \pi/4$). Therefore, we have 
 \begin{subequations}
     \begin{eqnarray}
   c_0 = B_{\rm cc}^{\rm pole},
 \label{Scconsts1}
 \end{eqnarray}      
   \begin{eqnarray}
   c_1 = \frac{B_{\rm cc}^{\rm eq}- c_0 }{u_{\rm cc}^{\rm eq}} ,
 \label{Scconsts2}
 \end{eqnarray}  
  \begin{eqnarray}
   c_2 = \frac{B_{\rm cc}^{\rm mid}- c_0 - c_1 u_{\rm cc}^{\rm mid} }{u_{\rm cc}^{\rm mid} \left( u_{\rm cc}^{\rm mid} - u_{\rm cc}^{\rm eq} \right)} .
 \label{Scconsts3}
 \end{eqnarray}  
 \end{subequations}
 This polynomial approximation produces acceptable results, since it only induces a negligible error on the crust-core boundary. Since $D_{\rm p}$ is a function of proton and neutron densities ($D_{\rm p}=D_{\rm p}(\rho_{\rm p},\rho_{\rm n})$) only, $\frac{\rm d \it D_{\rm p}}{du}=0$ and hence (\ref{Scbound1a}) takes the following form 
   \begin{eqnarray}
 y(u) = D_{\rm p}^{\rm cc} \tilde{B}_{\rm cc}(u) + 4\pi \left[ \frac{\rho_{\rm p}^{\rm crust}}{\rho_{\rm p}^{\rm core}} \right]_{\rm cc} M_{\rm N} (u),
 \label{Scyfun}
 \end{eqnarray}  
 which relates $y(u)$ with $M_{\rm N}(u)$. 

The second boundary condition ensures that $B_\phi$ is continuous at the crust-core boundary. Using (\ref{ffun}) and (\ref{fuN}) we obtain
       \begin{eqnarray}
    f_{\rm sc} (u)  =  \left[H_{\rm c1}\right]_{\rm cc}\frac{f_{\rm N}(u)}{\tilde{B}_{\rm cc}(u)},
 \label{Scbound2}
 \end{eqnarray}  
   which relates the superconducting and normal matter $f$ functions. It is obvious that the independent functions are now two, instead of four.
   Even though (\ref{ScGS}) does not depend explicitly on the superconducting function $M_{\rm sc}$, this function is used to obtain the equilibrium configurations, as will be shown later. The expression for $M_{\rm sc}$ is found by solving (\ref{yfun})
     \begin{eqnarray}
   M_{\rm sc} = \frac{1}{4\pi} \left[y(u) - B D_{\rm p} \right],
 \label{ScMsc}
 \end{eqnarray}
 and then substituting $y$ from (\ref{Scyfun}).

 \section{Numerical method}
 
 We use the Hachisu self-consistent field (HSCF) method
 \citep{Hachisu1} as presented by \cite{ Lander2012, Lander2ndB} with
 new convergence criteria to specify the equilibrium. First, we
 demonstrate these criteria and compare them to the original ones used
 by \cite{Hachisu1}.  Then, we describe the non-dimensional units, the plan of the method and finally the numerical implementation. The numerical code used to obtain the following solutions is an extension of the code presented in \cite{MasterThesis}.
 
  \subsection{New convergence criteria} 
 
 In the style of the original HSCF method, we would require that in order  for a numerical solution to be considered as a converged solution, the quantities 
  \begin{eqnarray}  
     \Delta \tilde{\mu}_{\rm p} = \left| \frac{\tilde{\mu}_{\rm p\: max, new}-\tilde{\mu}_{\rm p\: max, old}}{\tilde{\mu}_{\rm p\: max, new}} \right| ,
 \label{ScConvchP}
 \end{eqnarray}
 \begin{eqnarray} 
     \Delta \tilde{\mu}_{\rm n} = \left|\frac{ \tilde{\mu}_{\rm n\: max, new}-\tilde{\mu}_{\rm n\: max, old}}{ \tilde{\mu}_{\rm n\: max, new}-\tilde{\mu}}\right| ,
 \label{ScConvchN}
 \end{eqnarray}
 \begin{eqnarray}
     \Delta C_{\rm p} = \left| \frac{ C_{\rm p\:, new}-C_{\rm p, old}}{C_{\rm p\:, new}} \right|,
 \label{ScConvCP}
 \end{eqnarray}
  \begin{eqnarray} 
     \Delta C_{\rm dif} = \left|\frac{ C_{\rm dif\:, new}-C_{\rm dif, old}}{ C_{\rm dif\:, new}} \right|,
 \label{ScConvCdif}
 \end{eqnarray}
 should all have become less than a specified value (usually of order $10^{-6}$). Here, subscripts $new$ and $old$ denote consecutive iterative steps. However, these criteria proved to be adequate only for relatively simple models, which converge to high accuracy after a small number of iterations, such as  those presented in \cite{Hachisu1} and \cite{TomimuraEriguchi}.
 
In the case of entrainment, the system of equations is more complex, requiring a larger number of iterations to achieve convergence. In order to ensure that all quantities relevant for the solution have converged, we introduce three new convergence quantities related to the proton density $\rho_{\rm p}$, the neutron density $\rho_{\rm n}$ and the magnetic function $u$. The above physical quantities are   fundamental, in the sense that all other quantities are functions of them. Motivated by the definition of the usual standard deviation, we introduce the following normalized error measures
 \begin{eqnarray}
   \sigma_{\rho_{\rm p}} = \sqrt{ \frac{ \displaystyle{ \sum_{i=1}^{KDIV} \sum_{j=1}^{NDIV}  \left( \rho_{\rm p \:\, \textit{i,j} \:\rm new} - \rho_{\rm p \:\, \textit{i,j} \:\rm old}  \right)^2 } }{  NDIV \cdot KDIV}  } ,
 \label{Apconvrhop}
 \end{eqnarray}
  \begin{eqnarray}
   \sigma_{\rho_{\rm n}} = \sqrt{\frac{\displaystyle{ \sum_{i=1}^{KDIV} \sum_{j=1}^{NDIV}  \left( \rho_{\rm n \:\, \textit{i,j} \:\rm new} - \rho_{\rm n \:\, \textit{i,j} \:\rm old}  \right)^2 } }{NDIV \cdot KDIV}},
 \label{Apconvrhon}
 \end{eqnarray}
  \begin{eqnarray}
   \sigma_u = \sqrt{\frac{ \displaystyle{\sum_{i=1}^{KDIV} \sum_{j=1}^{NDIV}  \left( u_{i,j \:\rm new} - u_{i,j\:\rm old}  \right)^2 }}{NDIV \cdot KDIV}  },
 \label{Apconvu}
 \end{eqnarray}
where $NDIV$and $KDIV$ are the number of equidistant grid points in
the $r$ and $\mu:=\cos\theta$ directions respectively, while subscripts $i$, $j$ specify the grid point. All of the above quantities should converge to zero. The first two are sensitive to the convergence of the fluid properties, while the last one is sensitive to the convergence of the magnetic field.

When the magnetic field is not too strong, the density-dependent quantities (\ref{Apconvrhop}) and (\ref{Apconvrhon}) converge much faster than (\ref{Apconvu}) which is related to the magnetic field. In such cases, we decouple the matter fields from the magnetic field when 
(\ref{Apconvrhop}) and (\ref{Apconvrhon}) have  either reaches a plateau or have become sufficiently small and continue iterating only the magnetic field equation, until (\ref{Apconvu}) also reach a plateau or
becomes sufficiently small. The decoupling also prevents the growth of inherent numerical inaccuracies due to the finite precision of the method. We note  that (\ref{Apconvrhop}),(\ref{Apconvrhon}) and (\ref{Apconvu})
are global error indicators and are effectively averaged over the whole numerical  grid. In contrast, when using the original HSCF-style criteria 
(\ref{ScConvchP})-(\ref{ScConvCdif}), two of these quantities  are
local quantities and do not show a monotonous variation during
convergence -- see Section \ref{polent}. Using the new criteria (\ref{Apconvrhop}),(\ref{Apconvrhon}) and (\ref{Apconvu})
it is more straightforward to decouple the iteration of the magnetic field (provided the magnetic field is weak) from the iteration of the fluid quantities, once the latter have become of sufficient accuracy.

      \subsection{Non-dimensional Units}
In order to solve the equations numerically, we choose a system of non-dimensional units, which is based on the maximum density $\rho_{\rm max}$, the gravitational constant $G$ and the (proton) equatorial radius $r^{\rm p}_{\rm eq}$. In this system, the length, mass and time units are
  \begin{eqnarray}
     \left[ L \right] = r_{\rm eq}^{\rm p},
 \label{ScLenght}
 \end{eqnarray}  
  \begin{eqnarray}
    \left[ \mathcal{M} \right] \color{black} = \left( r_{\rm eq}^{\rm p} \right)^3 \rho_{\rm max},
 \label{ScMass}
 \end{eqnarray}
  \begin{eqnarray}
     \left[ T \right] = \frac{1}{\sqrt{4 \pi G \rho_{\rm max}}}.
 \label{ScTime}
 \end{eqnarray} 
 In Appendix \ref{APdimquant} we derive the dimensionless form of
 various quantities which are shown with `` $\hat{}$ ". Using the dimensionless units the form of the equations does not change. Dividing the proton chemical potential with the maximum value $\tilde{\mu}_{\rm max}$ and using (\ref{Scchpotdef1}), (\ref{Scchpotdef2}) as well as the definitions for the dimensionless densities (\ref{ScdimRhop}) and (\ref{ScdimRhon}), one obtains   
  \begin{eqnarray}
 \hat{\rho}_{\rm p} = x_{\rm p}(0)   \left( \frac{\tilde{\mu_{\rm p}}}{\tilde{\mu}_{\rm p\: max}} \right)^{N_{\rm p}} ,  
 \label{ScrhoPdef}
 \end{eqnarray} 
 and similarly
   \begin{eqnarray}
 \hat{\rho}_{\rm n}=\left(1- x_{\rm p}(0) \right)  \left( \frac{\tilde{\mu_{\rm n}}}{\tilde{\mu}_{\rm n\: max}} \right)^{N_{\rm p}}. 
 \label{ScrhoNdef}
 \end{eqnarray} 
 where $x_{\rm p}(0)$ is the proton fraction ($x_{\rm p}:=\rho_{\rm p} / \rho$) at the center of the star.

     \subsection{Plan of the method} \label{Convmain2}
      
 We specify the polytropic indices $N_{\rm p}$, $N_{\rm n}$ and the ratio of polar to equatorial radii for protons $\hat{r}^{\rm p}_{\rm pol}/\hat{r}^{\rm p}_{\rm eq}$, which is simply equal to $\hat{r}^{\rm p}_{\rm pol}$, since the dimensionless value of $\hat{r}^{\rm p}_{\rm eq}=1$. The equatorial radius of neutrons $\hat{r}^{\rm n}_{\rm eq}$ is determined through the definition of the crust-core surface (\ref{ccsurf}) and its dimensionless form is the ratio $\hat{r}^{\rm n}_{\rm eq}/\hat{r}^{\rm p}_{\rm eq}$. The functions $M_{\rm N}(u)$ and $f_{\rm N}(u)$ and the superconductivity parameter $h_{\rm c}$, as well as the central proton fraction $x_{\rm p}(0)$ and entrainment constant $\delta^\star$ are specified. For numerical reasons, we usually need to specify an under-relaxation parameter $\omega<1,$ to achieve convergence. The proton chemical potential vanishes at the surface of the star, while the neutron chemical potential vanishes at the crust-core surface. Using (\ref{ScIntP}) and (\ref{ScIntDif}) we evaluate $\hat{\tilde{\mu}}_{\rm p}$ at $\hat{r}_{\rm eq}^{\rm p}$, $\hat{r}_{\rm pol}^{\rm p}$ (where $\hat{\tilde{\mu}}_{\rm p}(\hat{r}^{\rm p}_{\rm eq},0) \equiv \hat{\tilde{\mu}}_{\rm p}(1,0)=0$, $\hat{\tilde{\mu}}_{\rm p}(\hat{r}^{\rm p}_{\rm pol},1)=0$ respectively) and $\hat{\tilde{\mu}}_{\rm n}$ at $\hat{r}^{\rm n}_{\rm eq}$ (where $\hat{\tilde{\mu}}_{\rm n}(\hat{r}^{\rm n}_{\rm eq},0)=0$) and we derive the equations for the  integration constants  $\hat{C}_{\rm p}$ and $\hat{C}_{\rm dif}$
  \begin{eqnarray} 
    \hat{C}_{\rm p} = \hat{\Phi}_{\rm g}(1,0) - \hat{M}(1,0),
 \label{ScCpEval}
 \end{eqnarray} 
  \begin{eqnarray}   
    \hat{C}_{\rm dif} = \hat{\tilde{\mu}}_{\rm p} (\hat{r}^{\rm n}_{\rm eq},0) - \hat{M}(\hat{r}^{\rm n}_{\rm eq},1) +\hat{N}(\hat{r}^{\rm n}_{\rm eq},1) .
 \label{ScCdifEval}
 \end{eqnarray} 
  The main iteration algorithm is:
  
\vskip0.5cm
  \noindent \underline{Main iteration}
 \begin{enumerate}
   \item Assign initial values  $\hat{\rho}_{\rm p}=1$, $\hat{\rho}_{\rm n}=1$ and $\hat{u}=1$.
   \item Compute $\hat{\Phi}_{\rm g}$ from (\ref{ScPhipot}).
   \item Calculate $\Pi$ from $\hat{u}$ (\ref{Pifun}).
   \item Compute intermediate $\hat{u}$ from (\ref{ScU1}) (i.e. $\hat{u}_{\rm int}$). Before evaluating, divide by $\Pi_{\rm max}$ and multiply again after integration.
   \item Employ under-relaxation to find the new $\hat{u}$         \begin{eqnarray}  
     \hat{u}_{\rm new}= (1-\omega) \hat{u}_{\rm int} +\omega \hat{u}_{\rm old},
 \label{underrelax}
 \end{eqnarray}
   \item Evaluate the proton integration constant $\hat{C}_{\rm p}$ from (\ref{ScCpEval}).
  \item Use the  proton integral equation (\ref{ScIntP}) to evaluate $\hat{\tilde{\mu}}_{\rm p}$.
  \item Evaluate the difference integration constant $\hat{C}_{\rm dif}$ from (\ref{ScCdifEval}).
  \item Using the difference integral equation (\ref{ScIntDif}), compute $\hat{\tilde{\mu}}_{\rm n}$.
  \item Compute new proton and neutron densities from (\ref{ScrhoPdef}) and (\ref{ScrhoNdef}).
  \item Return to first step and use for the next iteration the new values of $\hat{\rho}_{\rm p}$, $\hat{\rho}_{\rm n}$ and $\hat{u}$, until $\sigma_{\rho_{\rm p}}$, $\sigma_{\rho_{\rm n}}$ converge.
  \item Calculate $\Pi$ from $\hat{u}$ (\ref{Pifun}). (In the case of decoupling, hereafter we use fixed $\rho_{\rm p}$ and $\rho_{\rm n}$ values.)
  \item Compute intermediate $\hat{u}$ from (\ref{ScU1}). Before evaluating, divide by $\Pi_{\rm max}$ and multiply again after integration.
  \item Employ under-relaxation to find the new $\hat{u}$.
  \item Return to step (xii) until $\sigma_{u}$ has converged.

 \end{enumerate}
 We use the aforementioned scheme to solve the system of equations numerically, employing a two-dimensional grid in spherical polar coordinates, with points equidistant
in $r$ and $\mu:=\cos \theta$. Solving the integral equation for the gravitational
potential is already well known \citep{Hachisu1}. The equation for obtaining
$u$ as well as its numerical implementation are given  in Appendix \ref{GSimpelment}.

      \subsection{Virial test}
   
The scalar virial theorem derivation for the two-fluid model is similar to the single fluid  \citep[see][]{Chandra,Collins} and is given by 
    \begin{eqnarray}
  \frac{1}{2}\frac{\rm d \it ^2 I}{ \rm d \it t^2} = 2T + \mathcal{E}_{\rm
mag}  +W + 3 \frac{U_{\rm n}}{N_{\rm n}} + 3 \frac{U_{\rm p}}{N_{\rm p}}+
\mathcal{E}_{ \mbf{F}_{\rm n}} ,
 \label{ScVirial}
 \end{eqnarray}
 where $I$ is the moment of inertia, $T$ is the kinetic energy, $\mathcal{E}_{\rm
mag}$ is the magnetic energy given by
 \begin{eqnarray}
\mathcal{E}_{\rm mag} = \int_{\rm star} \mbf{r} \cdot \mbf{F}_{\rm
mag} \: dV ,
 \label{ScMagEn}
 \end{eqnarray}  
 $W$ is the gravitational energy of the system, $U_{\rm n}$ and $U_{\rm p}$
is the internal energy of neutrons and protons respectively and finally $\mathcal{E}_{\rm
\mbf{F}_n}$ is the energy due to the presence of entrainment given by
 \begin{eqnarray}
\mathcal{E}_{\rm \mbf{F}_n} = \int_{\rm  star} \mbf{r} \cdot \mbf{F}_{\rm
n} \: dV ,
 \label{ScEntrEn}
 \end{eqnarray}
 which, hereafter, we will refer to as the "entrainment energy".  
The form of the last equation is analogous to (\ref{ScMagEn}) and given
by the derivation of the scalar virial theorem \citep{Collins}. 
 
Since the moment of inertia needs to be constant with respect to time
for stationary solutions, the right side of the virial theorem
vanishes. Notice that we consider only nonrotating
models. Numerically, the virial theorem will differ from zero and so we construct the virial test quantity,
   \begin{eqnarray}
    VC = \frac{\left| \mathcal{E}_{\rm mag} +W + 3 \frac{U_{\rm n}}{N_{\rm
n}} + 3 \frac{U_{\rm p}}{N_{\rm p}} +\mathcal{E}_{\rm \mbf{F}_n} \right|}{\left|
W \right| } ,
 \label{ScVC}
 \end{eqnarray}
 which provides a test for the global convergence of the algorithm. Notice that for magnetized stars, this is a meaningful test only when the magnetic energy is a measurable fraction of the total energy.

 \section{Results} 
 
\subsection{Assumptions} 
 
 In order to obtain a numerical solution of an equilibrium model, we have to specify the functions and the various parameters governing the system of equations. For the normal matter magnetic functions $f_{\rm N}(u)$ and $M_{\rm N}(u)$ we have chosen
   \begin{eqnarray}
    f_{\rm N}(u)= a \left( u - u_{\rm int} \right)^{\zeta +1} \Theta \left( u- u_{\rm int} \right),
 \label{fNdefin}
 \end{eqnarray} 
  \begin{eqnarray}
    M_{\rm N}(u)= \kappa u,
 \label{MNdefin}
 \end{eqnarray} 
 where $a$, $\zeta$ and $\kappa$ are constants, $u_{\rm int}$ corresponds to the value of the largest constant-$u$ line that closes inside the star (i.e. the $u$ value on the equator) and $\Theta(u)$ is the Heaviside step function. We consider purely poloidal field models. The parameter $a$ determines the strength of the toroidal part of the magnetic field while the value of $\kappa$ is related to the strength of both the poloidal and toroidal part of the magnetic field in the mixed field case. In the specific models shown below, we omitted the  toroidal
component, since its contribution is typically very small compared to the
poloidal field, see \citep{LanderJones, Lander2ndB}. We choose  $x_{\rm p}(0)=0.15$, $h_{\rm c}=0.1$ and polytropic indices $N_{\rm n}=1$, $N_{\rm p}=1$. The mass of the neutron star is chosen as $ \mathcal{M}=1.64\,\rm M_\odot
$ for all configurations studied here. 

 Our numerical results are converted from non-dimensional units to Gaussian-CGS units using $\rho_{\rm max}=10^{15}\,\rm g\,cm^{-3}$ and $r^{\rm p}_{\rm eq}=15\,\rm km$. The value of the gravitational constant  is $G=6.673\times 10^{-8} \, \rm cm^{3}\,g^{-1}\,s^{-2}$.
The numerical grid consisted of $480\times 480$ points, incorporating one quadrant (using equatorial symmetry and axisymmetry) and an underrelaxation parameter $\omega=0.02$ was  necessary for achieving convergence. The initial guesses for all the grid points are  $\hat\rho_{\rm p}=1$, $\hat\rho_{\rm n}=1$ and $\hat{u}=1$.

  \begin{table*}
 \centering
\caption{Properties of three models with entrainment
(first two lines) and without (third line). The values of the convergence quantities $\log\sigma_{\rho_{\rm
n}}$, $\log\sigma_{\rho_{\rm p}}$ are shown at the last fully-coupled iteration while $\log\sigma_u$ is shown at the end of all iterations. }
 \scriptsize
\begin{tabular}{p{1.9cm}p{0.8cm} p{1.1cm}p{1.0cm}p{1.0cm}p{1.8cm}p{0.8cm}p{0.8cm}p{0.8cm}p{0.6cm}p{1.0cm}
} 
\hline
$\rm Model$ & $\delta_\star$ & $\mathcal{E}_{\rm mag}/\left|W\right| $ &
$\mathcal{E}_{\rm \mbf{F}_n}/\left|W\right|$ & $\left| W \right|$ & 
 $B_{\rm pole} \, (10^{15}\,\rm G) $  & $\log\sigma_u$ & $\log\sigma_{\rho_{\rm p}}$  &$\log\sigma_{\rho_{\rm
n}}$& $\kappa$ & $\rm Virial\; test$ \\
\hline
$\rm with \ entrainment$ & 0.8 & 3.33E-05  & 4.66E-07 & 6.12E-02  &  $1.16$ & -6.93 & -6.65 & -6.18 & 0.0320 &  1.48E-05  \\ [-0.3ex]
$\rm with \ entrainment$ & 0.9 & 1.37E-05 & 3.42E-07 & 6.11E-02  & $1.21$ & -6.88 & -5.96 & -5.71 & 0.0280 & 5.32E-06  \\ [-0.3ex]
$\rm {no \ entrainment} $ & 1.0 & 1.27E-05  & -  & 6.11E-02  &  $1.34$  & -6.58 &
-7.75 & -8.27 & 0.0278 & 3.53E-06   \\ 
\hline
\end{tabular}
\label{table1}
\end{table*}

    \begin{figure*}
    \includegraphics[scale=0.28]{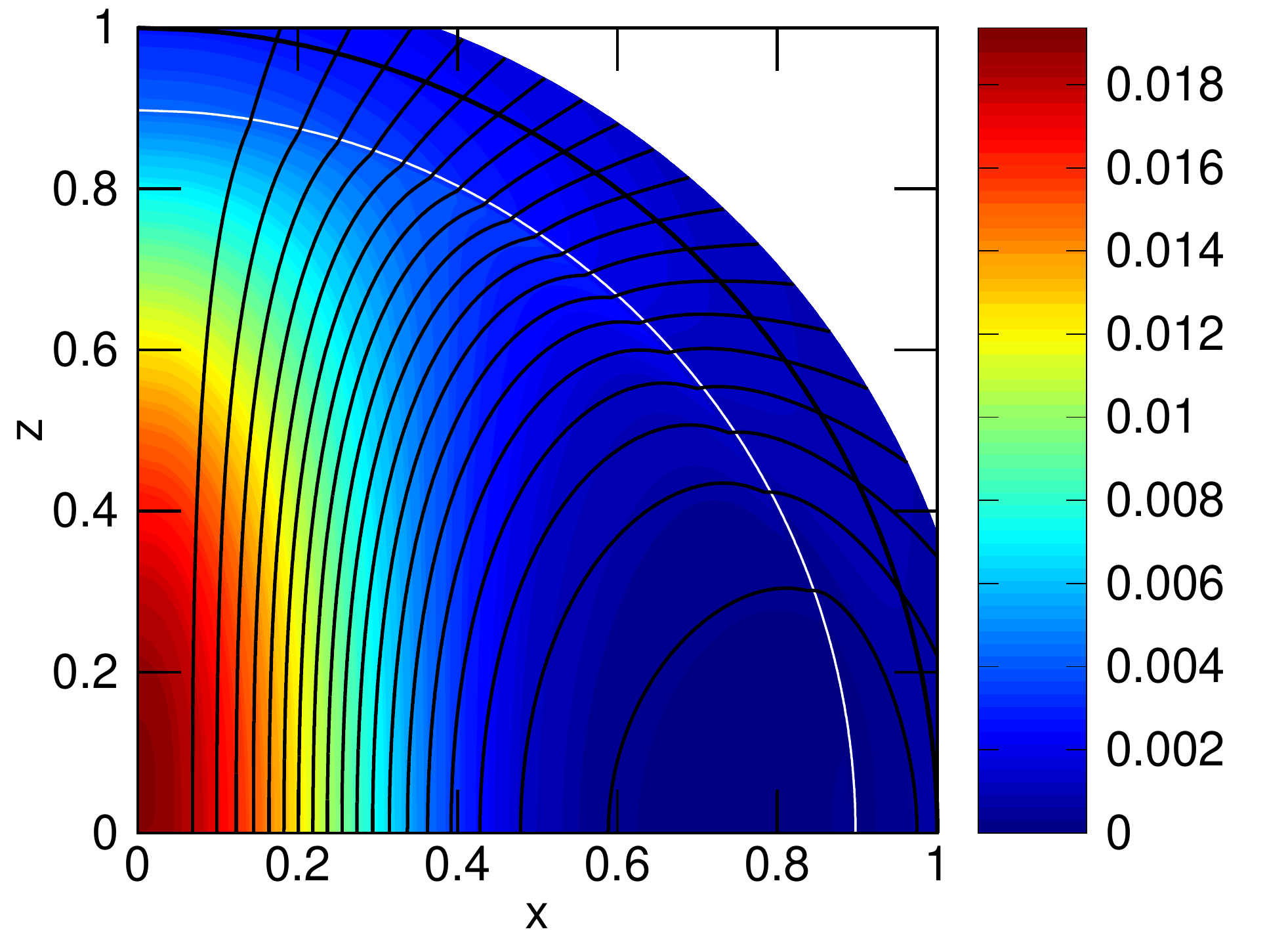}
    \includegraphics[scale=0.28]{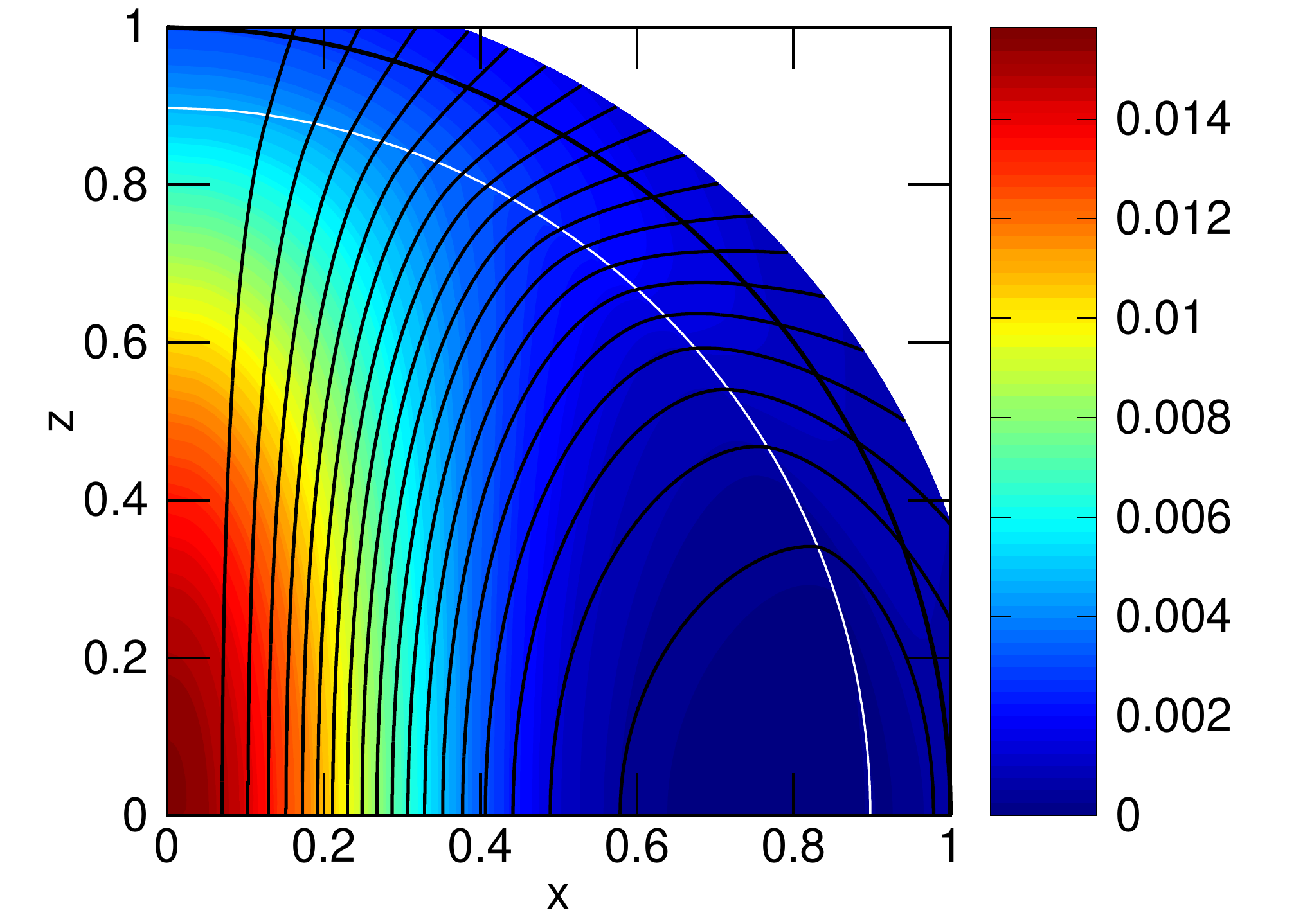}
     \includegraphics[scale=0.28]{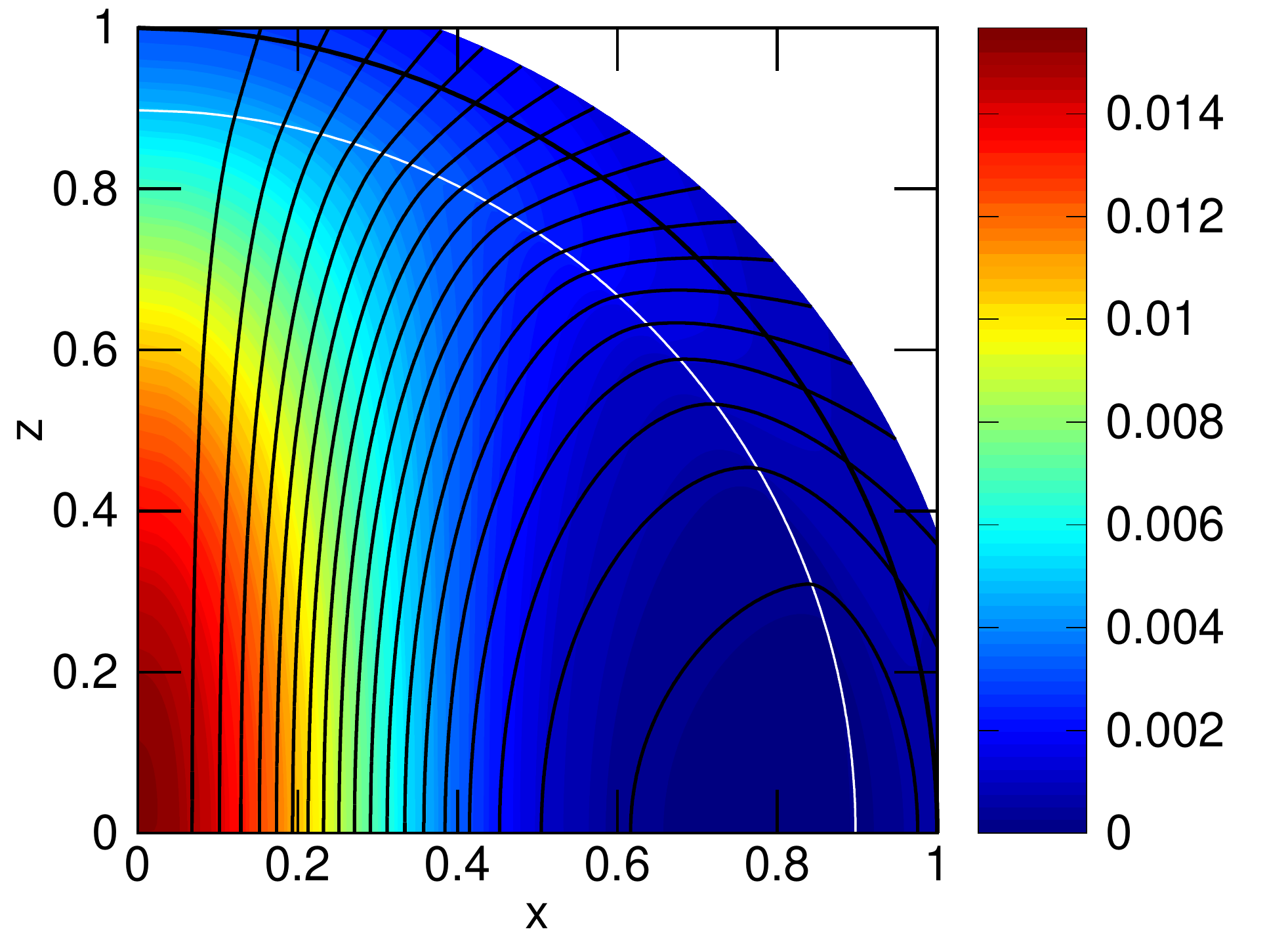}
   \caption{Poloidal magnetic field lines (contours of $u$) for the magnetized two-fluid type-II superconducting
neutron stars of Table \ref{table1}. The left-most and middle model have entrainment constant of
$\delta_\star=0.8, 0.9$ respectively. The right-most is without entrainment.
The black circle represents
the surface of the star while the white circle represents the crust-core boundary.
The color scale shows magnitude of the magnetic field in dimensionless units. The three cases differ mainly by the presence of kinks in the contours of $u$ at the crust-core boundary. }
   \label{figuBpol}
       \end{figure*}
       
 \subsection{The poloidal field configuration with entrainment}
 \label{polent}
 
  We focus on three different models with magnetic field strength of order $10^{15}$G at the pole, constructed with entrainment constant of $\delta_\star=0.8, 0.9$ and 1.0 (the latter value corresponds to the entrainment-free case). Since $\delta_\star$ is related to the proton density, we chose these representing values to obtain the equilibria \citep{Chamel}. Table \ref{table1} summarizes the main properties of the models.  The energy related to the entrainment is only a small fraction compared to the magnetic energy. The magnetic energy for the entrainment and entrainment-free cases is comparable for equilibria with similar magnetic fields.

\begin{figure}
    \hspace*{1cm}
    \includegraphics[scale=0.4]{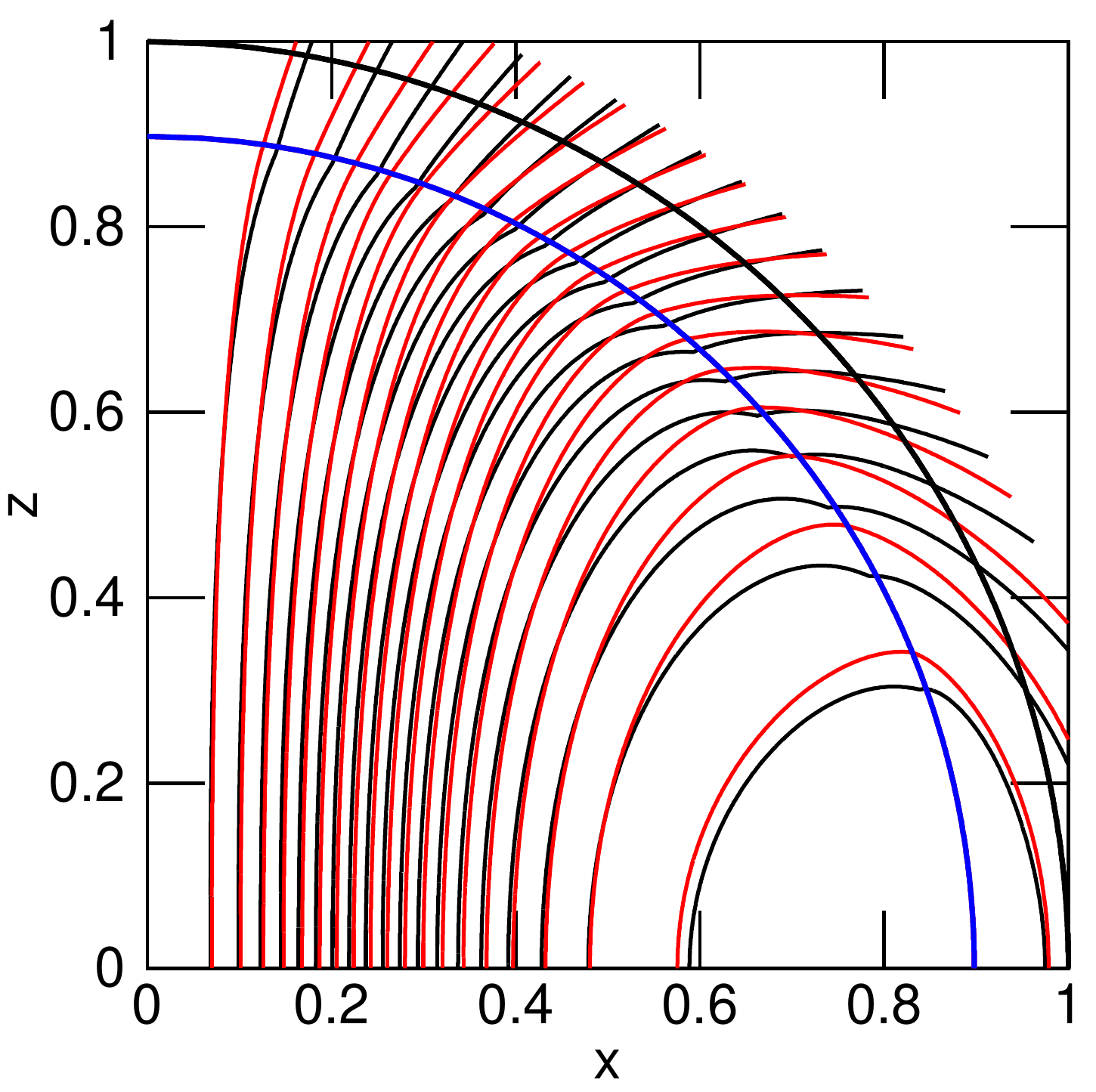}
   \caption{Comparison of magnetic field lines (contours of $u$) for  the
models with entrainment parameter $\delta_\star=0.8$ (black lines) and entrainment-free
model (red lines). The blue line represents the crust-core boundary.  }
   \label{ulinescomp}
       \end{figure}

 Fig.\ref{figuBpol}. shows the magnetic field lines (actually, contours of the magnetic function $u$) for the above three models. The main qualitative  difference among these models is the presence of kinks in the contours of $u$ at the crust-core boundary, which are especially evident in the $\delta_\star=0.8$ model. These kinks emerge because of the finite precision of the numerical treatment of the crust-core boundary. Matching the right-hand side values of \ref{ScGS} and \ref{ScGsnormal} on the crust-core boundary is less accurate than in the entrainment-free case. This happens because the type-II superconducting Grad-Shafranov equation (\ref{ScGS}) obtains a much more complicated form (due to non-constant entrainment parameter $\varepsilon_\star$) than that in the entrainment-free case presented in \cite{Lander2ndB}. Fig. \ref{ulinescomp} compares directly the $u$-contours between the $\delta_\star=0.8$ model and the entrainment-free case. The largest differences occur near the crust-core  boundary and in the crust region.

\begin{figure*}
              \begin{center}  
               \includegraphics[scale=0.4]{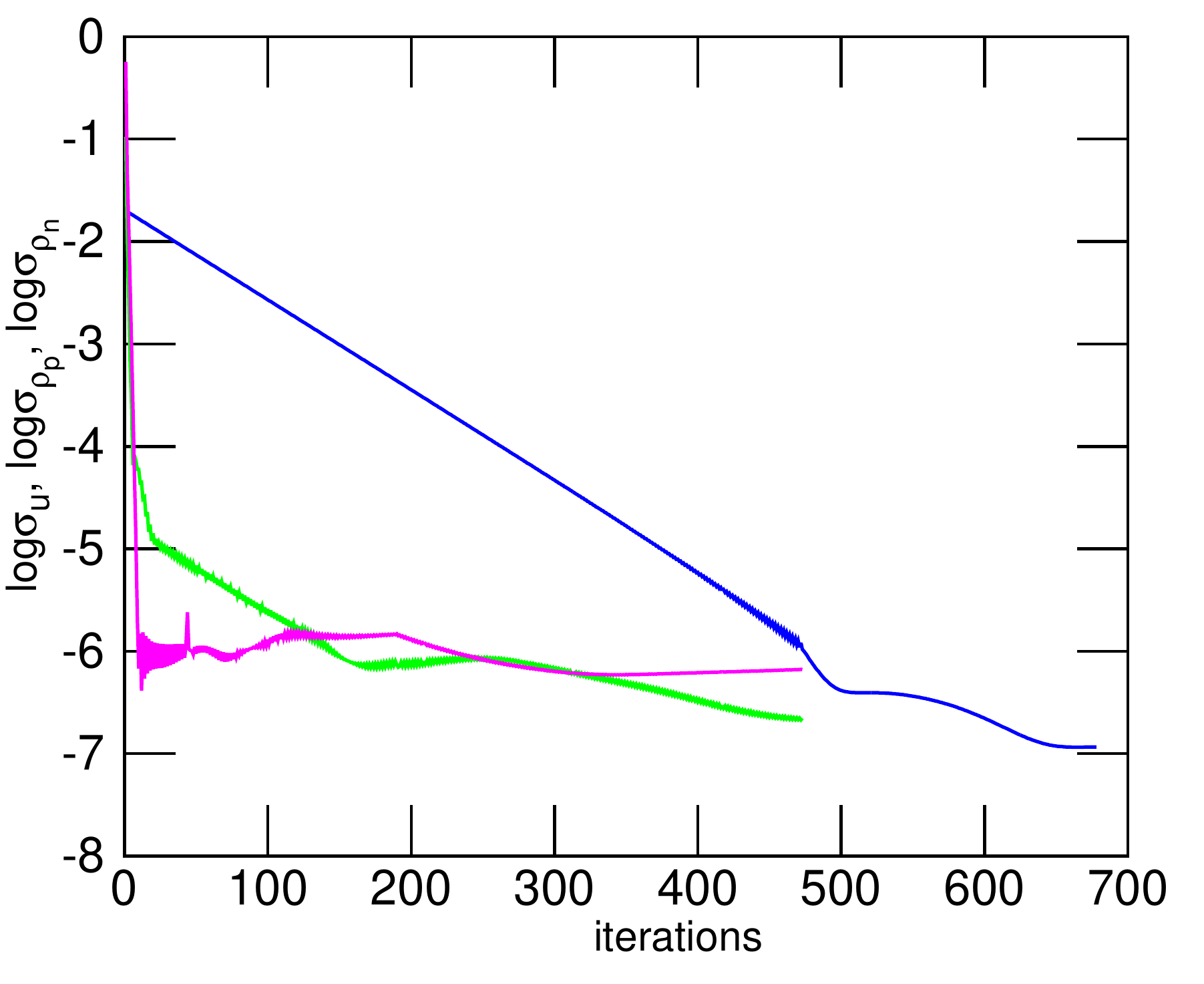} 
                \includegraphics[scale=0.4]{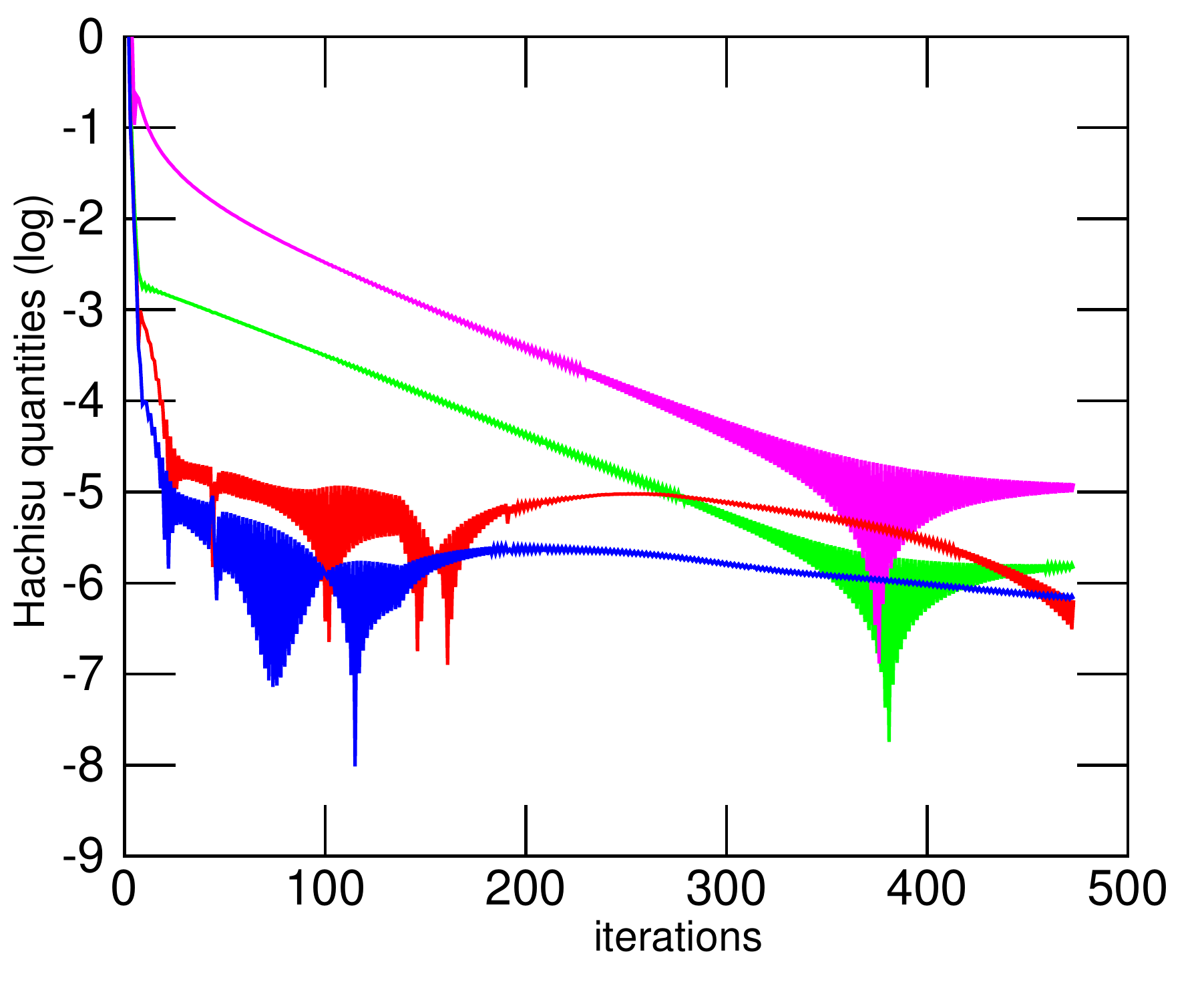} 
               \end{center}
                  \caption{ \textit{Left panel: }The new convergence test quantities, $\sigma_u$ (blue), $\sigma_{\rho_{\rm n}}$ (magenta) and $\sigma_{\rho_{\rm p}}$ (green)  as a function of iteration number for the case $\delta_\star=0.8$. The density-related test quantities reach a plateau before 500 iterations. At that point, we decouple the iteration of the magnetic field, in order to improve its accuracy. It takes about 200 additional iterations for the magnetic field test quantity (blue line) to reach a plateau. \textit{Right panel:} The Hachisu-style convergence test quantities, $\Delta \tilde{\mu}_{\rm n}$ (blue), $\Delta C_{\rm dif}$ (magenta), $\Delta C_{\rm p}$ (green),  and $\Delta \tilde{\mu}_{\rm p}$ (red). Their non-monotonous behaviour can terminate the iteration process too early.   }
                   \label{conv}       
               \end{figure*}

The low value of the magnetic energy with respect to fluid energies in the above models, implies that after the fluid quantities have numerically converged to their most accurate values, one could consider them as fixed and continue the iterations for  the magnetic field only, in order to improve its numerical convergence. The convergence test quantities as a function of the iteration number are shown in Fig. \ref{conv} (left panel) for the model with $\delta_\star=0.8$. After nearly 500 iterations, the accuracy of the fluid quantities does not improve anymore and we continue iterating only the magnetic field equation. It takes about 200 additional iterations
for the magnetic field test quantity (blue line) to reach a plateau.  \textit{In contrast, if one would continue to update the fluid quantities in the iteration process, this would only accumulate numerical truncation errors, which would not allow the magnetic field to converge in a satisfactory manner.} 

 For comparison, we provide a similar graph of the Hachisu-style convergence test quantities in Fig. \ref{conv} (right panel). It is obvious that these quantities vary significantly during iterations, which might result in erroneously terminating the iterations earlier than desired.
  

 \subsection{Stronger poloidal field results}
 
Here we discuss models with a stronger magnetic field (3 to 4 times larger at the pole than the models in the previous Section). We focus on two particular models, one with entrainment  $\delta_\star=0.8$ and one without entrainment. The main properties for both equilibria
are shown in Table \ref{table2}. In both
models, the magnetic energy and the entrainment energy are higher when compared
to the results of Table \ref{table1}. 

\begin{figure}
    \includegraphics[scale=0.4]{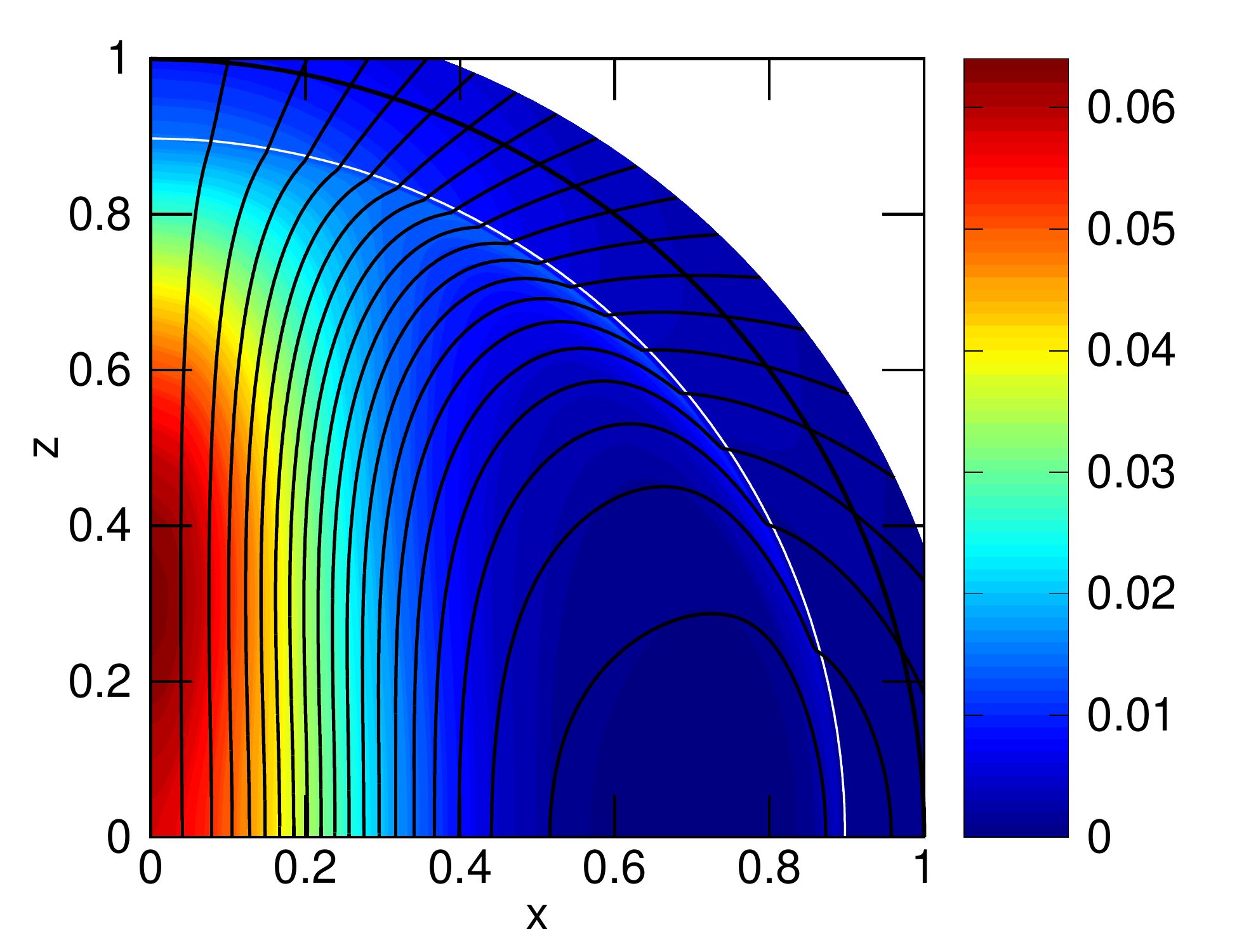}
   \caption{Poloidal field for a magnetized two fluid type-II superconducting neutron star with entrainment ($\delta_\star=0.8$). The black lines represent the u contours. The black circular line represents the surface of the star while the white line represents the crust-core boundary. We show the magnitude of the magnetic field with color. The magnetic field on the pole is $3.73\times 10^{15}\,G$. Here, the $u$ kinks on the crust core boundary are much more obvious than in the weaker field case. }
   \label{uBpolverystrong}
       \end{figure}

In both cases, the maximum value of the magnetic field in the converged solution does not appear at the center of the star, but is displaced along the z-axis to a point between $0.3\,R_{eq}$ and $0.4\,R_{eq}$  (Fig. \ref{uBpolverystrong}). This off-center maximum occurs when $B_{\rm pole}$ exceed roughly $3 \times 10^{15}\,G$. In addition, in these models with stronger magnetic field, the kinks in the $u$ contours on the crust-core boundary are much more apparent.  

Another aspect of the stronger field is that the convergence error indicators obtain a plateau at somewhat larger values than for the models in the previous Section. This implies that for models with stronger magnetic fields one would need better resolution to achieve the same accuracy.

         \begin{table*}
 \centering
\caption{Results for the stronger purely poloidal field configuration with entrainment and without. The values of the convergence quantities $\log\sigma_{\rho_{\rm
n}}$, $\log\sigma_{\rho_{\rm p}}$ are shown at the last fully-coupled iteration while $\log\sigma_u$ is shown at the end of all iterations. }
 \scriptsize
\begin{tabular}{p{1.9cm}p{0.8cm} p{1.1cm}p{1.0cm}p{1.0cm}p{0.8cm}p{1.8cm}p{0.8cm}p{0.8cm}p{0.8cm}p{0.6cm}p{1.0cm} } 
\hline
$\rm Model$ & $\delta_\star$ & $\mathcal{E}_{\rm mag}/\left|W\right| $ & $\mathcal{E}_{\rm \mbf{F}_n}/\left|W\right|$ & $\left| W \right|$ & $ \mathcal{M} $ &  $B_{\rm pole} \, (10^{15}\,\rm G) $  & $\log\sigma_u$ & $\log\sigma_{\rho_{\rm p}}$  &$\log\sigma_{\rho_{\rm n}}$& $\kappa$ & $\rm Virial\; test$ \\
\hline
$\rm with \ entrainment$ & 0.8 & 9.16E-05  & 4.65E-06 & 6.11E-02  & 0.968 & 3.73 & -5.90 & -5.96 & -5.63 & 0.0490 &  3.06E-05  \\ [-0.3ex]
$\rm no \ entrainment$ & 1.0 & 4.53E-05 & - & 6.12E-02  &  0.968 & 5.16  & -6.11 & -7.14 & -7.75 & 0.0420 & 4.31E-06  \\ 
\hline
\end{tabular}
\label{table2}
\end{table*}

 \section{Discussion} 

We constructed Newtonian equilibrium models for magnetized,
axisymetric neutron stars taking, into account  type-II
superconductivity of protons, superfluidity of neutrons and entrainment. We
employed the MHD equations presented in \cite{SCMHD}, a set of equations
accounting for the forces related to fluxtubes and vortices. As of
today, this is the most ``complete" set of type-II superconducting MHD equations in the Newtonian framework. The models presented here are derived from these equations by neglecting the effect of rotation on the structure of the star. Thus, the numerical models presented here are ``complete" in the sense that they employ all terms discussed in \cite{SCMHD} for non-rotating fluids in equilibrium. As pointed out in the introduction this paper provides an additional step towards our understanding of neutron star equilibria. The physics of entrainment is likely be of  importance in the case of fast rotating stars.

The poloidal field is qualitatively similar to the entrainment-free cases. The
energy due to entrainment is only a very small fraction of the total
energy -- about an order of magnitude less than the magnetic energy. The new convergence test criteria we introduce, allow for a more accurate  magnetic field configuration to be obtained. The main difference between models with and without entrainment and magnetic field at the pole less than $\sim 3 \times 10^{15}$G is structure of magnetic field lines along the crust-core boundary. In the entrainment case, kinks are present that do not appear in the entrainment-free case. Though, as previously mentionted, these kinks are of numerical nature. For  stronger field magnetic fields, these kinks are stronger while the magnetic field maximum appears off-center, along the symmetry axis.

It is possible to include rotation in our models, although even the simplest case, rigid rotation, requires a much more complicated form of equations (\ref{ScEulLanP}) and (\ref{ScEulLanN}) as various terms related to vorticity emerge. This constitutes a future step for this model to be achieved as well as including realistic equations of state rather than polytropes. Then, instead of approximating $\delta_\star$ as a constant, a relation between $\delta_\star$ and $\rho,$ as shown in \cite{Chamel}, can be employed as well. Furthermore, a more detailed description of the different physical properties of the inner and outer parts of the crust could be used as well. In the case of strong crustal entrainment as discussed in \cite{Chamel2} different configurations might arise. Finally, it is possible to model a magnetosphere by extending the magnetic function $f_N$ in the exterior of the neutron star, as shown in  \cite{magnetosphere}.

\section*{Acknowledgments}

We thank Nils Andersson, Konstantinos Glampedakis and Kyriaki Dionysopoulou for their helpful comments and discussions on this project.
Partial support comes from ``NewCompStar'', COST Action MP1304.
\bibliographystyle{mn2e_new}

\appendix

\section[]{Mathematical DERIVATIONS}

\subsection{Derivation of the Grad-Shafranov equation} \label{APgsderiv}

In order to derive the Grad-Shafranov equation for a type-II
superconductor we substitute $\mbf{H}_{\rm c1}=H_{\rm
  c1}\hat{\mbf{B}}$ in (\ref{ScMagForce}) and use (\ref{Dp}) to
  find that:
   \begin{eqnarray} 
   \mbf{F}_{\rm mag} &=& -\frac{1}{4\pi} \left[\mbf{B}\times \left(H_{\rm c1} \bm{\nabla} \times \hat{\mbf{B}} \right) \right. \nonumber \\ 
 &&  \left. + \mbf{B} \times \left( \bm{\nabla}H_{\rm c1} \times \hat{\mbf{B}} \right)    + \rho_{\rm p} \bm{\nabla} \left( B D_{\rm p} \right)   \right. \Big{]} .
  \label{ScMagForce1}
  \end{eqnarray}
   Following \citet{Lander2ndB}, we define a unit current $\hat{\mbf{j}}$ as
    \begin{eqnarray}
    \hat{\mbf{j}}:=\bm{\nabla}\times\hat{\mbf{B}} = \frac{1}{\varpi}\bm{\nabla}\left(\varpi \hat{B} \right) \times \mbf{e}_{\phi} + \hat{j}_{\phi}\mbf{e}_{\phi} .
  \label{Unitcurrent1}
  \end{eqnarray}
  Using (\ref{Unitcurrent1}) in (\ref{ScMagForce1}) and using vector calculus identities we obtain
    \begin{eqnarray} 
  -4\pi\,\mbf{F}_{\rm mag} &=& \rho_{\rm p}\bm{\nabla} \left(B D_{\rm p} \right) - \frac{1}{\varpi^2} \left( \bm{\nabla}\left( \varpi \hat{B}_{\phi} H_{\rm c1} \right)\times \bm{\nabla}u \right) \nonumber \\
    & &+\frac{1}{\varpi}B_\phi \bm{\nabla}\left(\varpi \hat{B}_\phi H_{\rm c1} \right)\nonumber \\ 
    & & + \frac{1}{\varpi}\left( \frac{\bm{\nabla}H_{\rm c1}\cdot \bm{\nabla}u  }{\varpi B}-H_{\rm c1} \hat{j}_\phi \right) \bm{\nabla}u. 
  \label{ScMagForce2}
  \end{eqnarray}
  Since for the magnetic force, $\mbf{F}_{\rm mag}=\rho_{\rm p}\bm{\nabla}M_{\rm sc}$ (see Eq. \ref{ScnablaMeqF}) the right side of the aforementioned equation should not contain any components in the $\phi$-direction. Thus, the toroidal part of (\ref{ScMagForce2}) should be zero
\begin{eqnarray}
 \bm{\nabla}\left(  \varpi \hat{B}_{\phi} H_{\rm c1} \right)\times \bm{\nabla}u=0.
\label{torpartzero}
\end{eqnarray} 
Setting now $f_{\rm sc}:= \varpi \hat{B}_{\phi} H_{\rm c1}$ we obtain that $f_{\rm sc}$ is
a function of $u$ (i.e. $f_{\rm sc}=f_{\rm sc}(u)$). We also set $y:=4\pi M_{\rm sc}+BD_{\rm p}$
and from (\ref{ScMagForce2}) we obtain  \begin{eqnarray} 
 -\bm{\nabla}y &= \frac{B_\phi}{\rho_{\rm p}\varpi}\frac{\rm d \it f_{\rm sc}}{\rm d \it u} \bm{\nabla}u + \frac{1}{\varpi} \left( \frac{\bm{\nabla}H_{\rm c1}\cdot \bm{\nabla}u  }{\varpi B \rho_{\rm p}}-\frac{H_{\rm c1}}{\rho_{\rm p}} \hat{j}_\phi \right) \bm{\nabla}u.
  \label{ScMagForce3}
  \end{eqnarray}
It is evident if we take the cross product of both parts of (\ref{ScMagForce3}) with $\bm{\nabla}u$ that $y=y(u)$. Substituting $\hat{B}_\phi$ from the definition of $f(u)$ we obtain

  \begin{eqnarray} 
 \frac{\rm d \it y}{\rm d\rm u} &= -\frac{B_\phi f_{\rm sc}}{\rho_{\rm p}\varpi H_{\rm c1}}\frac{\rm d \it f_{\rm sc}}{\rm d \it u} -\frac{\bm{\nabla}H_{\rm c1}\cdot \bm{\nabla}u  }{\varpi^2 B \rho_{\rm p}}+\frac{H_{\rm c1}}{\varpi \rho_{\rm p}} \hat{j}_\phi.
  \label{ScMagForce4}
  \end{eqnarray} 

 Using now the definition (\ref{Unitcurrent1}) of the unit current $\hat{j}$ and (\ref{ScBgen}) we obtain the relation between $\hat{j}_\phi$ and $u$
  \begin{eqnarray} 
   \hat{j}_\phi=-\frac{1}{\varpi B}\left[\Delta_\star u-\frac{1}{B} \bm{\nabla}B \cdot \bm{\nabla} u\right].
  \label{jphiu}
  \end{eqnarray} 
  Substituting the aforementioned equation in (\ref{ScMagForce4}), we obtain the type-II superconducting Grad-Shafranov equation, similar to the one presented in \cite{Lander2ndB}:

   \begin{eqnarray} 
    \Delta_\star u = \frac{\bm{\nabla} \Pi \cdot \bm{\nabla} u}{\Pi} -\varpi^2 \rho_{\rm p} \Pi \frac{\rm d \it y}{\rm d \it u}-\Pi^2f_{\rm sc}\frac{\rm d \it f_{\rm sc}}{\rm d \it u},
  \label{ScMagForce5}
  \end{eqnarray}
  
 where $\Pi$ is defined through (\ref{Pifun}).

\subsection{Entrainment quantities}

The explicit relations for $D_{\rm p}$, $D_{\rm n}$ as functions of $\rho_{\rm p}$ and $\rho_{\rm n}$ are
 \begin{eqnarray}
  D_{\rm p}:= \frac{\partial H_{\rm c1}}{\partial \rho_{\rm p}}=h_{\rm c}\,\frac{ \rho_{\rm p}^2\left( \delta_\star -1 \right)^2 + 2\rho_{\rm p}\rho_{\rm n}\left(\delta_\star-1 \right)\delta_\star +\rho_{\rm n}^2\delta_\star}{\left( \rho_{\rm p}\left(\delta_\star -1 \right)+\rho_{\rm n}\delta_\star \right)^2},
 \label{Dpexpl}
 \end{eqnarray}   
  \begin{eqnarray}
  D_{\rm n}:= \frac{\partial H_{\rm c1}}{\partial \rho_{\rm n}}=-h_{\rm c}\,\frac{ \rho_{\rm p}^2\left( \delta_\star -1 \right)^2 }{\left( \rho_{\rm p}\left(\delta_\star -1 \right)+\rho_{\rm n}\delta_\star \right)^2}. 
 \label{Dnexpl}
 \end{eqnarray}  
It is obvious that when $\delta_\star=1$ (i.e. $\varepsilon_\star=1$) $D_{\rm p}=1$ and $D_{\rm n}=0$.

\subsection{The boundary conditions}\label{APboundcond}

 The first boundary condition (\ref{Scbound1a}) dictates that $\left[B D_{\rm p}\right]_{\rm cc}$ is a function of $u$. Since $D_{\rm p}$ is a function of the neutron and proton densities only, it holds that
   \begin{eqnarray}
     \left[\frac{\rm d \it \left(BD_{\rm p} \right)}{\rm d \it u}\right]_{\rm cc}=\left[D_{\rm p}\frac{\rm d \it B}{\rm d \it u}\right]_{\rm cc}.
 \label{Dpdu}
 \end{eqnarray}  

\section{Numerics}
\subsection{The numerical implementation of the Grad-Shafranov equation}\label{GSimpelment}

 The Grad-Shafranov equation can be written as a Poisson equation using (\ref{deltaoper}) and hence it can be inverted and solved for $u$ as
 \begin{eqnarray}
  u(r,\mu) = \frac{\varpi}{4 \pi \sin{\phi} } \int_{\rm all\:\: space} \frac{F(\mbf{r'})}{\left| \mbf{r'}-\mbf{r} \right|} \sin{\phi'} dV', 
 \label{ScU}
 \end{eqnarray}  
 where $F(\mbf{r})$ is defined through 
     \begin{eqnarray}
 F   =
               \left\{ \begin{array}{lr}
           - \frac{\bm{\nabla} \Pi \cdot \bm{\nabla} u }{\varpi \Pi} +\varpi \rho_{\rm p} \Pi \frac{\rm d \it y}{\rm d \it u} + \frac{\Pi^2}{\varpi}f \frac{\rm d \it f}{\rm d \it u} ,             & \text{core}  \\
                   4\pi \varpi \rho_{\rm p} \frac{\rm d \it M_{\rm N}}{du}+\frac{f_{\rm N}}{\varpi} \frac{\rm d \it f_{\rm N}}{\rm d \it u} ,      & \text{crust}  \\ 
                      0,   & \text{exterior}
                         \end{array} \right. 
 \label{ScFtotal}
 \end{eqnarray}      
 (note that the minus sign by inverting
Poisson's equation has been absorbed in $F$). The primed quantities describe source points while non-primed points
 are where $u$ is calculated. Expanding  $\frac{1}{\left|
     \mbf{r'}-\mbf{r} \right|}$ in ($r$, $\mu$) coordinates and due to
 axisymmetry, (\ref{ScU1}) takes the following form:
       \begin{eqnarray}
  u(r,\mu) &=& \varpi \int_{r'=0}^{+\infty} dr'\int_{\mu'=0}^{1} d \mu'
   \left[ \sum_{n=1}^{+\infty} \tilde{f}_{2n-1}(r',r) \right. \nonumber 
  \\ 
  & & \frac{1}{2n \left( 2n-1 \right)}  P_{2n-1}^1 \left( \mu \right) P_{2n-1}^1 \left( \mu' \right) \Bigg] F(r',\mu'), \nonumber\\ 
 \label{ScU1}
 \end{eqnarray} 
 where $P_{l}^{m}(\mu)$ are the associated Legendre polynomials and  $\tilde{f}_n(r',r)$ is the radial part of the expansion, given by
   \begin{eqnarray}
 \tilde{f}_{n}(r',r)=\left\{  \begin{array}{lr}
   \frac{r'^n}{r^{n+1}} ,& r>r' \\
   \frac{r^n}{r'^{n+1}} ,& r<r'  \end{array}
        \right. .
 \label{f2n}
 \end{eqnarray}
We now discretize (\ref{ScU1}) in order to obtain a relation that can
be implemented numerically. Following the HSCF method \citep{Hachisu1}
we employ a 2D $r$ vs. $\mu$ grid with NDIV$\times$KDIV points in the
respective directions. The coordinate $\mu$ ranges between 0 and 1
while the radial coordinate $r$ ranges between 0 and $r_{\rm max}$ (we set
$\hat{r}_{\rm max}$=16/15 to cover the whole star and a small
  exterior region). The grid points  are
thus given by    \begin{eqnarray}
 r_j=r_{\rm max}\frac{(j-1)}{NDIV-1},
 \label{rdirection}
 \end{eqnarray}
and
   \begin{eqnarray}
 \mu_i=\frac{(i-1)}{KDIV-1}.
 \label{mudirection}
 \end{eqnarray} 
 We compute $n=LMAX$  terms of the Legendre and associated Legendre polynomials (here we choose $LMAX=16$). Using the same notation as before, integrating (\ref{ScU}) using Simpon's rule, over $\mu'$ yields
    \begin{eqnarray}
   U_{k,n}^{(1)} &=& \sum_{l=1}^{KDIV-2} \frac{1}{3(KDIV-1)}  \left( P_{2n-1}^1(\mu_l) F_{k,l}\right. \nonumber \\
 & & \left. + 4\, P_{2n-1}^1(\mu_{l+1}) F_{k,l+1} + P_{2n-1}^1(\mu_{l+2})F_{k,l+2} \right),\nonumber \\ 
 \label{ScUNumInt1}
 \end{eqnarray} 
while integration over $r'$ gives
      \begin{eqnarray}
   U_{n,j}^{(2)} &=& \sum_{k=1}^{NDIV-2} \frac{r_{\rm max}}{3(NDIV-1)}  \left( r^2_{k} \tilde{f}_{2n-1}(r_k,r_j) U_{k,n}^{(1)} \right. \nonumber \\
 & & \left. + 4\,r^2_{k+1}\tilde{f}_{2n-1}(r_{k+1},r_j) U_{k+1,n}^{(1)} \right. \nonumber \\
 & & \left.  + r^2_{k+2}\tilde{f}_{2n-1}(r_{k+2},r_j)U_{k+2,n}^{(1)} \right) .
 \label{ScUNumInt2}
 \end{eqnarray} 
 Then, $u$ is given by
   \begin{eqnarray}
  u_{i,j} = r_j \sqrt{1-\mu_i^2}\, \sum_{n=1}^{LMAX}\frac{1}{2n\left( 2n-1 \right)} U_{n,j}^{(2)} P_{2n-1}(\mu_{i}) ,
 \label{ScUNumInt2}
 \end{eqnarray} 
where $U_{k,n}^{(1)}$ and $U_{n,j}^{(2)}$ are intermediate integration quantities.
 
\subsection{Dimensionless quantities}\label{APdimquant}
 
 Here we list various quantities in their dimensionless form 
 \begin{eqnarray}
        \hat{r} = \frac{r}{r_{\rm eq}^{\rm p}},
 \label{ScdimR}
 \end{eqnarray} 
  \begin{eqnarray}
        \hat{\rho}_{\rm p} = \frac{\rho_{\rm p}}{\rho_{\rm max}},
 \label{ScdimRhop}
 \end{eqnarray} 
  \begin{eqnarray}
        \hat{\rho}_{\rm n} = \frac{\rho_{\rm n}}{\rho_{\rm max}},
 \label{ScdimRhon}
 \end{eqnarray} 
  \begin{eqnarray}
      \hat{\Phi}_{\rm g} = \frac{\Phi_{\rm g}}{4
      \pi G \left(r_{\rm eq}^{\rm p}\right)^2 \rho_{\rm max}},
 \label{ScdimPhig}
 \end{eqnarray} 
 \begin{eqnarray}
      \hat{C}_{\rm p} = \frac{C_{\rm p}}{ 4
      \pi G \left(r_{\rm eq}^{\rm p}\right)^2 \rho_{\rm max}},
 \label{ScdimCp}
 \end{eqnarray} 
    \begin{eqnarray}
      \hat{C}_{\rm dif} = \frac{C_{\rm dif}}{4
      \pi G \left(r_{\rm eq}^{\rm p}\right)^2 \rho_{\rm max}},
 \label{ScdimCdif}
 \end{eqnarray}  
  \begin{eqnarray}
    \hat{\tilde{\mu}}_{\rm p} = \frac{\tilde{\mu}_{\rm p}}{4
      \pi G \left(r_{\rm eq}^{\rm p}\right)^2 \rho_{\rm max}},
 \label{ScdimChemP}
 \end{eqnarray}  
 \begin{eqnarray}
    \hat{\tilde{\mu}}_{\rm n} = \frac{\tilde{\mu}_{\rm n}}{4
      \pi G \left(r_{\rm eq}^{\rm p}\right)^2 \rho_{\rm max}},
 \label{ScdimChemN}
 \end{eqnarray}  
 \begin{eqnarray}
    \hat{u} = \frac{u}{\sqrt{4
      \pi G} \left( r_{\rm eq}^{\rm p} \right)^3 \rho_{\rm max}} , 
 \label{ScdimU}
 \end{eqnarray}  
  \begin{eqnarray}
  {  \hat{\mathcal{M}} } =\frac{\mathcal{M}}{ \left( r_{\rm eq}^{\rm p} \right)^3 \rho_{\rm max}} , 
 \label{ScdimM}
 \end{eqnarray}  
  \begin{eqnarray}
    \hat{\kappa} =\frac{\kappa}{ \frac{\sqrt{4\pi G}}{r_{\rm eq}^{\rm p}}} ,
 \label{Scdimkappa}
 \end{eqnarray}  
  \begin{eqnarray}
    \hat{a} =\frac{a}{ \frac{1}{\sqrt{4 \pi G}\rho_{\rm max} \left(r_{\rm eq}^{\rm p}\right)^4 }} ,
 \label{Scdima}
 \end{eqnarray}  
 while for all quantities with energy units the dimensionless form is
 \begin{eqnarray}
    \hat{E} =\frac{E}{4 \pi  G  \left(r_{\rm eq}^{\rm p}\right)^5 \rho_{\rm max}} .
 \label{ScdimEn}
 \end{eqnarray}

\bsp

\label{lastpage} 

\end{document}